\documentclass{article}
\usepackage[T1]{fontenc} 
\usepackage[utf8]{inputenc} 
\usepackage{ismir,amsmath,cite,url}
\usepackage{graphicx}
\usepackage{booktabs}
\usepackage{color}
\usepackage{makecell}
\usepackage{caption} 
\usepackage{booktabs}
\usepackage{multirow}
\usepackage{rotating}
\usepackage{array} 
\usepackage{tabularray}
\usepackage{xcolor}
\usepackage{placeins}  

\usepackage[firstpage]{draftwatermark}
\definecolor{lightgray}{rgb}{0.9,0.9,0.9}
\definecolor{darkgray}{rgb}{0.4,0.4,0.4}
\SetWatermarkFontSize{12pt}
\SetWatermarkScale{1.1}
\SetWatermarkAngle{90}
\SetWatermarkHorCenter{202mm}
\SetWatermarkVerCenter{170mm}
\SetWatermarkColor{darkgray}
\SetWatermarkText{Late-Breaking / Demo Session Extended Abstract, ISMIR 2024 Conference}

\def\lbd{}	

\title{Do Captioning Metrics reflect \\ Music Semantic Alignment?}

\multauthor
{Jinwoo Lee\textsuperscript{1} \hspace{1cm} Kyogu Lee\textsuperscript{1,2,3}} { \\
 $^1$ Department of Intelligence and Information, Seoul National University\\
$^2$ Interdisciplinary Program in Artificial Intelligence, Seoul National University\\
$^3$ Artificial Intelligence Institute, Seoul National University\\
{\tt\small {\{e-{}-jw, kglee\}}@snu.ac.kr}
}

\def\authorname{Jinwoo Lee, and Kyogu Lee}

\begin{document}

\maketitle

\begin{abstract}
Music captioning has emerged as a promising task, fueled by the advent of advanced language generation models. However, the evaluation of music captioning relies heavily on traditional metrics such as BLEU, METEOR, and ROUGE which were developed for other domains, without proper justification for their use in this new field. We present cases where traditional metrics are vulnerable to syntactic changes, and show they do not correlate well with human judgments. By addressing these issues, we aim to emphasize the need for a critical reevaluation of how music captions are assessed.


\end{abstract}
\section{Introduction}\label{sec:introduction}
The advancement of music information retrieval (MIR) parallels significant developments in the music industry, particularly regarding tasks like music captioning \cite{muscaps, lpmusiccaps, gardner2023llark, li2023mert, enriching}.
Despite the promise of music captioning, the evaluation of generated captions poses significant challenges. Current metrics, borrowed from natural language processing tasks, do not adequately address the unique qualities of musical content. These traditional metrics, such as BLEU\cite{papineni2002bleu}, METEOR\cite{meteor}, and ROUGE\cite{lin2004rouge}, primarily rely on n-gram overlap, focusing on superficial similarities between generated and reference captions \cite{gardner2023llark}. This approach inherently favors syntactic similarity over semantic meaning, which may result in misleading evaluations of caption quality. For example, a paraphrased caption may receive a substantially lower score, even though it conveys the same core idea as the original.

Moreover, these metrics fail to effectively capture the semantics of musical contexts, such as genre, instruments, and mood. This can result in evaluations that may overlook critical aspects of the music being described, leading to poor alignment with how humans perceive and interpret musical content. This work aims to illustrate the limitations of these traditional metrics, and highlight instances where they score inaccurately high.




\section{Method} 

To evaluate the effectiveness of traditional metrics in music captioning, we conduct a human evaluation study aimed at correlating human judgments with the scores generated by these metrics. %


We use Amazon Mechanical Turk\cite{mturk} to recruit 50 participants for a listening test. From the evaluation set of the MusicCaps \cite{musiclm} dataset, we randomly sample 30 audio clips. Each audio sample is paired with three different types of captions:

\begin{itemize}
    \item Original: The ground truth caption associated with the audio in the dataset.
    \item Inference: The caption generated by the audio captioning model. For the inference, we use music captioning model by LP-MusicCaps \cite{lpmusiccaps}. 
    \item Paraphrased: A version of the original caption syntactically altered without changing the meaning.
       
\end{itemize}

Participants are asked to evaluate how accurately the captions describe the audio clips. To maintain the integrity of the evaluation and minimize potential bias, each participant is presented with only two of the three caption versions for each audio sample. This design helps prevent participants from recognizing that two of three captions are likely ground truth captions, while the other one is not.



\begin{table*}[h]
\centering
\caption{Examples of different caption types and their evaluation scores. We report BLEU (B) as their average.}
\begin{tblr}{
  row{1} = {c},
  row{2} = {c},
  cell{1}{1} = {r=2}{},
  cell{1}{3} = {c=4}{},
  cell{3}{2} = {c},
  cell{3}{3} = {c},
  cell{3}{4} = {c},
  cell{3}{5} = {c},
  cell{3}{6} = {c},
  cell{3}{7} = {c},
  cell{4}{2} = {c},
  cell{4}{3} = {c},
  cell{4}{4} = {c},
  cell{4}{5} = {c},
  cell{4}{6} = {c},
  cell{4}{7} = {c},
  cell{5}{2} = {c},
  cell{5}{3} = {c},
  cell{5}{4} = {c},
  cell{5}{5} = {c},
  cell{5}{6} = {c},
  cell{5}{7} = {c},
  cell{6}{2} = {c},
  cell{6}{3} = {c},
  cell{6}{4} = {c},
  cell{6}{5} = {c},
  cell{6}{6} = {c},
  cell{6}{7} = {c},
  cell{7}{2} = {c},
  cell{7}{3} = {c},
  cell{7}{4} = {c},
  cell{7}{5} = {c},
  cell{7}{6} = {c},
  cell{7}{7} = {c},
  cell{8}{2} = {c},
  cell{8}{3} = {c},
  cell{8}{4} = {c},
  cell{8}{5} = {c},
  cell{8}{6} = {c},
  cell{8}{7} = {c},
  hline{1,3,6,9} = {-}{},
}
Caption     & Human & N-gram Overlap &      &      &      & Embedding \\ 
& MOS &B    & M    & R    & S    & FENSE     \\
{\textit{\textcolor{gray}{Original}}\\Someone is playing a melody on an \textbf{e-guitar} with a \textbf{tremolo effect}.\\
This song may be playing at home practicing \textbf{guitar}.} & 4.60   & 1.00 & 1.00    & 1.00    & 1.00    & 1.00 \\
{\textit{\textcolor{gray}{Distorted (Semantic X, Syntactic O)}}\\
Someone is playing a melody on a \textbf{french horn} with a \textbf{very reverb}.\\
This song may be playing at home practicing \textbf{french horn}.}  & - & \textbf{0.61} & \textbf{0.38} & \textbf{0.73} & 0.33 & 0.56 \\
{\textit{\textcolor{gray}{Paraphrased~(Semantic O, Syntactic X)}}\\An \textbf{electric guitar} with \textbf{tremolo effect} plays a melody,\\ 
possibly during home practice.}   & 4.70   & 0.13  & 0.25 & 0.28 & \textbf{0.36} & \textbf{0.83}  \\

{\textit{\textcolor{gray}{Original}}\\This is a \textbf{yodeling} music piece. There is a \textbf{female} vocalist that is\\
singing \textbf{happily} in the lead. The melody is provided by \textbf{medium}\\
\textbf{and high pitch woodwinds}. In the background, the \textbf{bass line} is\\
played by an \textbf{upright bass} while the rhythm is provided by \\
an \textbf{acoustic} drum. The atmosphere is very \textbf{lively}. This piece\\
could be used in the soundtrack of a \textbf{comedy movie or}\\
\textbf{a children's show}.} & 4.64  & 1.00 & 1.00    & 1.00    & 1.00   & 1.00  \\
{\textit{\textcolor{gray}{Distorted (Semantic X, Syntactic O)}}\\This is  a \textbf{hard rock} piece. There is a \textbf{male} vocalist that is singing\\in \textbf{energetic} mood. The melody is provided by \textbf{low-pitch electric}\\\textbf{guitars}. Meanwhile, the \textbf{melody} is played by a \textbf{high-pitched violin}\\while the rhythm is provided by \textbf{electronic} drums. This piece could\\be used in the soundtrack of a \textbf{urgent action movie}.}                                    & -     & \textbf{0.43}           & \textbf{0.26} & \textbf{0.59} & \textbf{0.32} & 0.66      \\
{\textit{\textcolor{gray}{Paraphrased~(Semantic O, Syntactic X)}}\\A \textbf{lively} \textbf{yodeling} song with a \textbf{female} vocalist, \textbf{woodwinds}, \textbf{upright}\\\textbf{bass}, and \textbf{acoustic} drums, suitable for a \textbf{comedy or children's show}.}                                                                       & 4.32  & 0.04           & 0.15 & 0.29 & 0.30 & \textbf{0.72}     
\end{tblr}
\label{showcase}
\end{table*}

\begin{table*}[h]
\centering
\caption{Experimental results of correlation coefficient (Pearson's $r$) between evaluation metrics vs. human evaluation (MOS) for 20 entries in the evaluation set of MusicCaps dataset.}
\begin{tabular}{cccccccc}
\hline
\textbf{BLEU$_1$} & \textbf{BLEU$_2$} & \textbf{BLEU$_3$} & \textbf{BLEU$_4$} & \textbf{METEOR} & \textbf{ROUGE} & \textbf{SPICE} & \textbf{FENSE} \\ \hline
0.075 & 0.074 & 0.077 & 0.085 & 0.096 & 0.083 & 0.097 & 0.091 \\ \hline
\end{tabular}
\label{correlation_coefficient}
\end{table*}

\begin{figure}[h!]
\centering
\includegraphics[width=0.9\columnwidth]{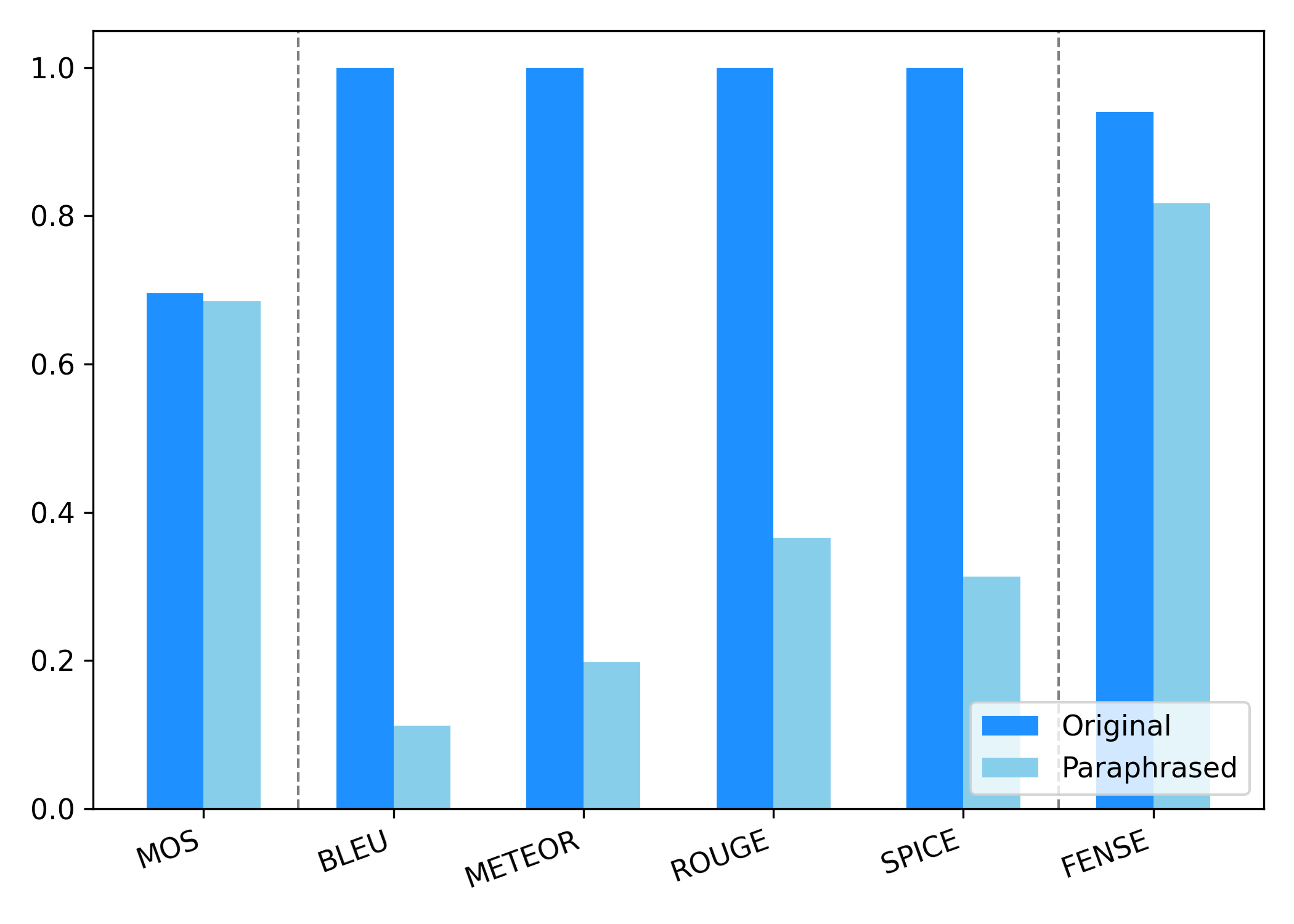}  
\caption{Experimental results on scores across caption type and evaluation metrics from MusicCaps evaluation set. MOS values are rescaled to [0, 1].}
\label{fig:syntax}
\end{figure}


\section{Results}


\subsection{Syntactic \& Semantic variations}
Following the collection of human evaluation scores, we report the scores of evaluation metrics (including the Mean Opinion Score (MOS)) for original and paraphrased captions, as shown in Figure~\ref{fig:syntax}. Notably, evaluation metrics except for FENSE show a significant decrease when comparing the original captions to their paraphrased versions. This indicates that these metrics may be overly sensitive to syntactic changes rather than semantic content. As illustrated in Table~\ref{showcase}, we also showcase examples of various caption types along with their corresponding evaluation scores. We also include distorted captions that are semantically altered while preserving their syntactic structure. Notably, most n-gram-based metrics tend to favor distorted captions over paraphrased captions.

\subsection{Correlation with human judgment}
To further validate our findings, we compute the correlation coefficient between the scores obtained from human evaluations and each of the evaluation metrics. As presented in Table~\ref{correlation_coefficient}, results indicate that most evaluation metrics, including FENSE\cite{fense}, do not show significant correlations with human evaluations.

\section{Conclusion}
We demonstrate existing metrics are overly sensitive to syntactic variations, and they lack alignment with actual human evaluations. Given these findings, we conclude that a more nuanced evaluation framework is necessary to address these challenges.


\bibliography{ref}

\end{document}